\documentclass[10pt,conference]{IEEEtran}

\IEEEoverridecommandlockouts
\usepackage{cite}
\usepackage{amsmath,amssymb,amsfonts}
\usepackage{pifont}
\usepackage{algorithmic}
\usepackage{graphicx}
\usepackage{textcomp}
\usepackage{xcolor}
\definecolor{LightGrey}{gray}{0.95}
\usepackage{setspace}
\usepackage{color, soul}
\usepackage{multirow}
\usepackage{longtable}
\usepackage{verbatim}
\usepackage{url} % typeset URL's reasonably
\usepackage{listings}
\usepackage{hyperref}
\usepackage{float}
\usepackage{xspace}
\usepackage{colortbl}
\usepackage{etoolbox}
\usepackage{svg}
\usepackage{placeins} 
\usepackage{multirow}

\usepackage{lineno,hyperref}
\modulolinenumbers[5]
\usepackage{booktabs}
\usepackage{adjustbox}
\usepackage{scrextend} % for footref (allows duplicate footnote references)
\usepackage{enumitem}
\usepackage[numbers]{natbib}
\AtBeginEnvironment{quote}{\itshape}

\def\BibTeX{{\rm B\kern-.05em{\sc i\kern-.025em b}\kern-.08em    T\kern-.1667em\lower.7ex\hbox{E}\kern-.125emX}}

\newcommand{\supplementary}{\href{https://doi.org/10.48420/c.6366063}{DOI: 10.48420/c.6366063}}

\newcommand{\POne}{\textcolor{blue}{P1}}\newcommand{\PTwo}{\textcolor{blue}{P2}}
\newcommand{\PThree}{\textcolor{blue}{P3}}\newcommand{\PFour}{\textcolor{blue}{P4}}
\newcommand{\PFive}{\textcolor{blue}{P5}}\newcommand{\PSix}{\textcolor{blue}{P6}}
\newcommand{\PSeven}{\textcolor{blue}{P7}}
\newcommand{\PEight}{\textcolor{blue}{P8}}\newcommand{\PNine}{\textcolor{blue}{P9}}
\newcommand{\PTen}{\textcolor{blue}{P10}}

\newcommand{\CSharp}{\href{https://docs.microsoft.com/en-us/dotnet/csharp/tour-of-csharp/}{C\#}}
\newcommand{\Java}{\href{https://www.oracle.com/java/}{Java}}
\newcommand{\Python}{\href{https://www.python.org}{Python}}
\newcommand{\RLang}{\href{https://www.r-project.org}{R}}
\newcommand{\DotNET}{\href{https://dotnet.microsoft.com/en-us/}{.NET}}
\newcommand{\Swift}{\href{https://www.swift.org}{Swift}}
\newcommand{\Typescript}{\href{https://www.typescriptlang.org}{Typescript}}
\newcommand{\Consolee}{\texttt{Console}}
\newcommand{\Double}{\texttt{Double}}
\newcommand{\Float}{\texttt{Float}}
\newcommand{\Integer}{\texttt{Integer}}

\newcommand{\Scanner}{\texttt{Scanner}}
\newcommand{\String}{\texttt{String}}
\newcommand{\charAt}{\texttt{.charAt}}
\newcommand{\EndsWith}{\texttt{.endsWith}}
\newcommand{\equals}{\texttt{.equals}}
\newcommand{\length}{\texttt{.length}}
\newcommand{\println}{\texttt{.println}}
\newcommand{\startsWith}{\texttt{.startsWith}}
\newcommand{\substring}{\texttt{.substring}}
\newcommand{\While}{\texttt{while}}
\newcommand{\toString}{\texttt{toString}}
\newcommand{\IF}{\texttt{if}}
\newcommand{\RUN}{\texttt{Run()}}
\newcommand{\SystemIn}{\texttt{System.in}}
\newcommand{\return}{\texttt{return()}}

\newcommand{\GetChar}{\texttt{getchar}\xspace}

\newcommand{\AND}{\textit{``AND"}\xspace}
\newcommand{\regex}{\textit{``regex"}\xspace}
\newcommand{\TwoDimensional}{\textit{``two-dimensional array"}\xspace}
\newcommand{\foreach}{\textit{``foreach"}\xspace}

\newcommand{\Matcher}{\texttt{Matcher}}

\newcommand{\Main}{\texttt{main} method}
\newcommand{\For}{\texttt{for} loop}

\newcommand{\NVivo}{\href{https://www.qsrinternational.com/nvivo-qualitative-data-analysis-software/home/}{NVivo 12}}
\newcommand{\Dropbox}{\href{https://www.dropbox.com/}{Dropbox}}
\newcommand{\Geeks}{\href{https://www.geeksforgeeks.org}{GeeksforGeeks}}

\newcommand{\JavaInTwoSemestersCitation}{\cite{quentin2019java}}

\newcommand{\JTpoint}{\href{https://www.javatpoint.com}{JavaTpoint}}
\newcommand{\SO}{\href{https://stackoverflow.com}{Stack~Overflow}}
\newcommand{\StackOverflow}{\SO}
\newcommand{\TutPoint}{\href{https://www.tutorialspoint.com/index.htm}{TutorialPoint}}
\newcommand{\WThree}{\href{https://www.w3schools.com}{W3school}}
\newcommand{\Zoom}{\href{https://zoom.us}{Zoom}}
\newcommand{\JSixSevenOne}{``Java671"}
\newcommand{\BegBook}{\href{https://beginnersbook.com}{BeginnerBook}}
\newcommand{\Tutorial}{Tutorial}

\newcommand{\etal}[1]{#1 \emph{et al.\@\xspace}}

\newcommand{\ie}{i.e.\@\xspace}
\newcommand{\eg}{e.g.\@\xspace}

\newcommand\mycite[2][]{\citeauthor{#2}\ (\citeyear{#2})\ifx#1\undefined\else, #1\fi}

\begin{document}

% \font\myfont=cmr12 at 64pt
\title{Novice Programmers Strategies for Online Resource Use and Their Impact on Source Code\\
\thanks{This research was sponsored by Saudi Electronic University, College of Computing and Informatics, Kingdom of Saudi Arabia.}}

\author{\IEEEauthorblockN{Omar Alghamdi\IEEEauthorrefmark{1}\IEEEauthorrefmark{2}, Sarah Clinch\IEEEauthorrefmark{1}, Mohammad Alhamadi\IEEEauthorrefmark{1}, and  Caroline Jay\IEEEauthorrefmark{1}}
 \vspace{0.1em}
 \IEEEauthorblockA{\IEEEauthorrefmark{1} Department of Computer Science, University of Manchester, M13 9PL,United Kingdom\\
 \IEEEauthorblockA{\IEEEauthorrefmark{2}College of Computing and Informatics, Saudi Electronic University, Riyadh,6867, Saudi Arabia\\[2pt]
 Email: 
 \IEEEauthorrefmark{1}$\langle{}$firstname$\rangle{}$.$\langle{}$lastname$\rangle{}$@manchester.ac.uk, \IEEEauthorrefmark{2}oalghamdi@seu.edu.sa}}}

\maketitle

% For page numbering (i.e. in review)
%\thispagestyle{plain}
%\pagestyle{plain}

\begin{abstract}    
    Websites are frequently used by programmers to support the development process. This paper investigates programmer-Web interactions when coding, and combines observations of behaviour with assessments of the resulting source code. We report on an online observational study with ten undergraduate student programmers as they engaged in programming tasks of varying complexity. % Programmers' activities were video recorded, their written code outputs were collected, and each participated in an interview. 
    Screens were recorded of participants' activities, and each participated in an interview. Videos and interviews were thematically analysed. Novice programmers employed various strategies for seeking and utilising online knowledge. %, increasing coding %complexity.  
    The resulting source code was examined to determine the extent to which it met requirements and whether it contained errors.
    % The source code analysis revealed that the coding complexity increased the possibility of producing qualified code.
    The source code analysis revealed that coding with the websites involved more coding time and effort, but increased the possibility of producing correct code.
    % On the other hand, other instances of source code were impacted and were either not working or not compliant with the requirements. 
    However, coding with websites also introduced instances of either incorrect or non-executable source code.
    % It can be assumed that more complexity caused by using the websites helps in writing qualified code; however, coding with the websites introduces risks of producing code that is not qualified.
    %It can be assumed that coding using the websites does not always guarantee an outcome that is correct and executable.

\end{abstract}

% Count: 191
% Progress: edited, require multiple edits.
% language: checked grammars, words, and sentences.  

\section{Introduction}

% What the intro should have:
    % Start with: programming is complex. Searching the websites is common. Code is necessary. Reusing online code is predominate. The investigations of the code issues presented online is largely covered. The impact of using websites on the produced code was limited. This paper aim to..

% AIM:
    % Impacts of Web on SRC: achieved by fetching activities, and analyse SRC. The link is new. 
    % Also, link SRC with code issues. 
 
% Paragraphs:
    % Web usage
    % Activites: search and reuse
    % Code issues
    % Arguments: could be in above paragrpahs?
    % Paper aim and sections. 
 
% Web usage
The internet has made information more accessible than ever before, with significant implications for how people learn and practice programming. 
Website usage is viewed as an integral part of the software engineering process \cite{yang2017stack}. Developers sometimes spend more time searching for information online than coding, assisting their learning and clarifying terms \cite{brandt2009two}. 
%Technical question and answer websites such as \StackOverflow{} are particularly popular; as of June 2022, \StackOverflow{} constitutes $22$ million questions and $34$ million answers, a clear indication of scale and popularity \cite{stack}. 

% Activities: search and reuse
    % Use of the Web lead to acquire reside knowledge to advance programmers. (?)
Searching is considered a vital process that is conducted pervasively to augment programmers' knowledge, even for familiar issues \cite{latozaMaintainingMentalModels2006, singerExaminationSoftwareEngineering2010, liWhatHelpDevelopers2013, sadowski2015developers}. In addition to providing information, searching the websites also aids in code understanding and error correction
\cite{gallardo-valenciaWhatKindsDevelopment2011, brandt2009two}. Programmers repeatedly access previously visited Web pages \cite{gaoExploringProgrammersAPI2020}. However, searching can affect programmers' productivity, as using the websites for simple tasks can take more time than coding from scratch, and searches may not always be successful \cite{wangCharacterizingDeveloperBehavior2017}. It can be difficult for programmers to locate and assess online code snippets \cite{escobar-avilaSurveyOnlineLearning2019}, which are often of low-quality \cite{xiaWhatDevelopersSearch2017}. Students, in particular, experience challenges, given their lack of vocabulary to assist their searches \cite{wong-aitkenItDependsWhether2022}.

% Activities: reuse
Seeking knowledge online is usually coupled with utilising code \cite{fuchsMonitoringStudentsMobile2014}, providing a structural template and supporting similar functionalities \cite{kim2004ethnographic}. Copying code from online sources is commonplace during development, especially from the website \SO \cite{brandt2009two, umarjiArchetypalInternetScaleSource2008, huckaSoftwareSearchNot2018, xiaWhatDevelopersSearch2017}. There is evidence that online queries are focused primarily on code reuse \cite{hora2021}
%With the importance of knowledge during coding, the explorations of programmers' online seeking and utilisation activities are under-explored lacking in-depth explorations. 
and that increased complexity in source code leads to greater online information seeking \cite{astromskisPatternsDevelopersBehaviour2017}. To date, there has been little exploration of how seeking and utilising online knowledge affects the production of source code. %, providing detailed information about how the programmers use online knowledge and reach their code with exploring possible difficulties. 

%  Code issues & SRC
Online code is often of poor quality or unsuited to the task at hand, making it challenging to use effectively. Prior research shows that code snippets on \StackOverflow{} contain problematic code such as security issues \cite{acar2016you, fischer2017stack, meng2018secure} and outdated statements \cite{ragkhitwetsagulToxicCodeSnippets2019}. %and failure to meet language-appropriate styling \cite{Bafatakis2019}. %Programmers' outputs could include code issues, given their possible exploitation of online content. Using \SO during coding introduces security issues to programmers' code \cite{acar2016you}. The propagation of the online code to the programmers' code is a possible outcome of seeking and utilising online content. To the best of our knowledge, this is the first paper to identify the extent of code problems to programmers' source code when coding with websites.
 
% OLD argument, but possible to consider:
    % In this paper, we argue that the quality of the presented online code and the mere present of the websites play a vital factor influencing the source code and produce deficient code. 

% Study method and RQs 
The current study documents how programmers use the websites to complete a series of coding tasks, and examines how these activities may affect the resulting source code. We screen recorded ten undergraduate students solving four programming tasks with varying levels of difficulty. We also collected the resulting source code for analysis and interviewed the programmers to understand in more depth why they used particular strategies and how these affected their code. %This experiment set out to capture in-depth observations of coding processes undertaken by programmers. Such observations provide the strategies employed using the resources (particularly online) to reach the required output and how these strategies reflect on the source code.  
Our research questions were as follows: 

 \begin{enumerate}[label=RQ\arabic*]
 \IEEEsetlabelwidth{12)}
    \item How do programmers use websites during programming?\label{RQ1}  
    \item What are the effects of websites use on the resulting code? \label{RQ2}
 \end{enumerate}
 
% Study results:
Our results show that programmers %prefer to seek and utilise online knowledge during coding. 
use various strategies to seek and utilise online knowledge. 
% These strategies produced complexity that could increase the programmers' coding time and efforts, but also increases the chances of producing qualified code.
These strategies may increase task completion time and effort but increases the chances of producing correct code. However, website use does not guarantee functional code -- in several instances participants still fail to produce code that executes and/or meets requirements. 
The results of this study shed the light in coding with the websites with a unique approach applying multiple inputs of data to the investigation, as well as in the assessment of the resulting code to discover problematic code propagation.

% Our hypotheses:
% \begin{itemize}
%     \item  The websites search during coding produces complexities such as unnecessary and not successful searches. 
%     \item  The adoption of online code cause complex issues with more time and efforts. 
%     \item  Searching websites and using online code results in source code that is either non-functional or non-compliant with the requirements. % you could separate it to two hypothesis. 

% % To include in the hypothesis:
% % -SRC success based on the strategies. 
% % Suggested HYPOTHESES:
% %     SRC: Searching and adopting online code does not guarantee producing working or compliant source code.
 
% % Notes about the hypothesis in Qualitative approach:
% % MY study is basically have Phenomenological type of qualitative. 
% % The hypothesis is a result of the inductive qualitative approach: based on observations.
% % You are not testing these hypothesis, but likely the hypothesis is the outcome of your research. 

% \end{itemize}

% NO NEED: I HAVE EMBDED ON THE DISCUSSION
%  Our contributions in this study is threefold: 
%     \begin{itemize}
%         \item  Determine what the relationship between the resources, specially the websites, and the process of coding
%         \item  Determine the followed strategies by the programmers to use the resources to code and produced the outcomes
%         \item  Investigate whether the purpose and timing of each strategies have an impact of the coding and the outcomes
%          \item Investigate the effectiveness of each strategies to reach the outcomes. 
%     \end{itemize}

\section{Related work}
Our work builds on research from software engineering that seeks to understand programmers' activities, and examines the resulting code to investigate possible implications.

\subsection{Understanding coding activities}

% Quantitative
In computer science education, \textit{learning analytics} 
%are used to understand novice programmers' coding practices. For example, 
such as Error Quotient~\cite{jadudFirstLookNovice2005,jadudMethodsToolsExploring2006} and Watwin Score~\cite{watsonPredictingPerformanceIntroductory2013} (both based on compilation behaviour and errors), NPSM~\cite{carterNormalizedProgrammingState2015a, carterBlendingMeasuresProgramming2017} (based on a more holistic set of IDE-captured measures), and RED~\cite{beckerNewMetricQuantify2016} (a measure of repeated errors), have been used to quantify programming behavior and predict outcomes. 
Outside of teaching, IDE instrumentation has been used to identify problem solving strategies \cite{robillardHowEffectiveDevelopers2004} and code/documentation interaction patterns \cite{astromskisPatternsDevelopersBehaviour2017, wangCharacterizingDeveloperBehavior2017} (\eg{} investigate, edit, validate \cite{schroerRecordingVisualisingUnderstanding2021}). %Those who engaged in systematic planning, investigation and recording of findings were most effective, solving problems more quickly \cite{robillardHowEffectiveDevelopers2004}. 

% Qualitative
Qualitative approaches are also used to characterise programmer behaviours \cite{arya2023programmers, Chatterjee2020, dedieuHowDevelopersSearch2022, escobar-avilaSurveyOnlineLearning2019, koExploratoryStudyHow2006, murphyDebuggingGoodBad2008, wangCharacterizingDeveloperBehavior2017, wong-aitkenItDependsWhether2022, ko2007information}. 
For example, \etal{Ko} analysed screen captures for experts engaging in software maintenance tasks, identifying three common activities -- code search, dependency following, and code collection \cite{koExploratoryStudyHow2006}. 
Combining IDE instrumentation with surveys and interviews has been used to understand professionals' debugging activities, with \texttt{printf} and breakpoints used in preference to more advanced IDE functionality \cite{bellerDichotomyDebuggingBehavior2018}; interviews have also been used to understanding student debugging \cite{murphyDebuggingGoodBad2008}. 
%Another study consider students' reasoning while coding in the wild and monitored ten students solving reverse engineering task, and  recorded and analyzed students' video and audio recordings \cite{kennedyQualitativeObservationsStudent2019}.  Their findings stress that creating a working code is not necessary mean that participants understand the employed features and concepts, and it mean that there are instances of trial and error error that caused accidental such success.
%In an Amazon Turk study, \cite{jacquesStudyingProgrammerBehaviour2021} analysed programmers' behaviours while solving two tasks. They found participants reading questions specifications, thinking about the problem, and understanding the code. These activities interfered with either typing or copying code without specifying the source. They also found that participants produced working code in the second task faster with fewer attempts than in the first task; they presume this might be because of the learning effect, familiarity, less searching, and less use of pasted code. 

% Intro to Web use: search and code cloning
Whilst the above focus on in-editor activity, this study seeks to better understand the usage of websites during coding activities. 
Such behaviors have previously been observed by \etal{Han}, who combined eye-tracking with IDE and browser logs to capture behaviour in Python %-based data quality assessment
tasks. In addition to quantifying code production, their study identifies patterns of external resource use (\eg{} searching, copy-paste) \cite{hanUnderstandingDataWorker2020}.

\subsubsection{Search}
%Prior research has considered search behaviour in light of information foraging theory \cite{koExploratoryStudyHow2006, chattopadhyayContextProgrammingInvestigation2018}. 
Programmers use their coworkers for information \cite{ko2007information}, and search both within their own codebases \cite{chattopadhyayContextProgrammingInvestigation2018, koExploratoryStudyHow2006, sadowski2015developers, sillitoManagingSoftwareChange2005, singerExaminationSoftwareEngineering2010, starke2009searching} and for online information (\eg{} through \href{http://www.google.com}{Google's Web search}) \cite{acar2016you, brandt2009two, chattopadhyayContextProgrammingInvestigation2018, hanUnderstandingDataWorker2020, koenzenCodeDuplicationReuse2020, liWhatHelpDevelopers2013, michaels2020empirical}. 
In this paper, we focus on the interactions when coding with websites, and our discussion of search therefore centers on this latter behavior. 
%
% Where the searches end up: Stack Overflow vs. tutorials
Google has been observed to dominate programming-related search, with the \StackOverflow{} Q\&A website most prominent in results \cite{acar2016you, escobar-avilaSurveyOnlineLearning2019, hora2021, wong-aitkenItDependsWhether2022, xiaWhatDevelopersSearch2017}. 
%Thus the majority of programmers' online search results in a visit to \StackOverflow{} \cite{baiExploringToolsStrategies2019, xiaWhatDevelopersSearch2017}, which is often used in preference to official documentation \cite{ragkhitwetsagulToxicCodeSnippets2019}. 
However, the behaviour of novices and students may differ, \eg{} using search to identify online tutorials in preference to \StackOverflow{} \cite{brandt2009two, koenzenCodeDuplicationReuse2020, bai2020graduate}. 

% When and why people search
Search engines are used to support code comprehension and reuse \cite{hora2021, maalejComprehensionProgramComprehension2014, xiaWhatDevelopersSearch2017, bai2020graduate}, debugging \cite{gallardo-valenciaWhatKindsDevelopment2011, hora2021, xiaWhatDevelopersSearch2017}, information acquisition and reference \cite{brandt2009two, gallardo-valenciaWhatKindsDevelopment2011, hora2021, xiaWhatDevelopersSearch2017}. 
Search frequency increases when tasks become complex \cite{astromskisPatternsDevelopersBehaviour2017, michaels2020empirical}, and both searches and their results are often consulted multiple times in a single session \cite{koenzenCodeDuplicationReuse2020, gaoExploringProgrammersAPI2020}. 
For example, \etal{Astromskis} found low rates of search among professionals ($6\%$ of sessions), but observed intense online consultation during some sessions with more complex code leading to greater use of the websites~\cite{astromskisPatternsDevelopersBehaviour2017}.

% Difficulties in searching
Searching websites are not always (immediately) successful. 
In observations, Wang noted that some participants did not click any of the search results, or that viewing a website was immediately followed by new search behavior~\cite{wangCharacterizingDeveloperBehavior2017}. %Taken together, these observations indicate that a non-trivial proportion of programmers' search activities are unsuccessful. %, and Wang notes that participants lacked search strategies even for very simple tasks. 
Other studies confirm that multiple queries may be needed to identify appropriate resources \cite{astromskisPatternsDevelopersBehaviour2017, gaoExploringProgrammersAPI2020, wong-aitkenItDependsWhether2022, xiaWhatDevelopersSearch2017}. 
This may be a result of difficulties representing code and symbols in queries; retrieval of unanswered questions; and retrieval of resources that are of poor quality, missing exemplars, or are difficult to understand \cite{escobar-avilaSurveyOnlineLearning2019, xiaWhatDevelopersSearch2017}. 
Effective search may be particularly problematic for novices, who lack strategies and vocabulary and have difficulty assessing the relevance of results \cite{Chatterjee2020, liWhatHelpDevelopers2013, nygrenTrackingStudentsInternet2017, wong-aitkenItDependsWhether2022}.
However, professionals also report challenges, particularly with regard to the volume of information retrieved \cite{dedieuHowDevelopersSearch2022}. 
Finally, context-switches arising from search may themselves be problematic, reducing productivity \cite{astromskisPatternsDevelopersBehaviour2017}. %, and there is some evidence that use of search engines and \StackOverflow{} may result in more queries than those who are limited to official documentation \cite{acar2016you}.

\subsubsection{Code cloning/reuse} 
A common objective for searching websites is the identification of code snippets for reuse  \cite{hora2021, fuchsMonitoringStudentsMobile2014, huckaSoftwareSearchNot2018, nygrenTrackingStudentsInternet2017, umarjiArchetypalInternetScaleSource2008, xiaWhatDevelopersSearch2017}. 
In one analysis of search queries, %that led users to five popular programming websites (StackOverflow{}, \WThree{}, \Geeks{}, \TutPoint{}, and programcreek), 
$46\%$ were seeking code for reuse~\cite{hora2021}. 
%(\eg{} ``javascript read line from file``) %as opposed to information seeking ($20\%$ general concepts, $9\%$ programming, $8\%$ databases), debugging ($8\%$) or tool support ($6\%$) \cite{hora2021}.

Studies indicate that both students \cite{bliksteinUsingLearningAnalytics2011, fuchsMonitoringStudentsMobile2014, meierBehaviourPatternsLearners2020, vihavainenHowNovicesTackle2014} and professionals \cite{astromskisPatternsDevelopersBehaviour2017, ciborowskaDetectingCharacterizingDeveloper2018, jacquesStudyingProgrammerBehaviour2021, kim2004ethnographic} engage in code cloning, although there is some evidence that students may adopt alternative approaches as they develop expertise~\cite{vihavainenHowNovicesTackle2014}. 
Professionals report using code clones as a tool to help refactor code, act as a structural template/example, and support forks/branches \cite{kim2004ethnographic, latozaMaintainingMentalModels2006}, 
and both groups engage in follow-on modification activity, \eg{}: complete or partial removal, correction and compilation cycles, modification and `beautification' to fit context, and addition of new code \cite{baiExploringToolsStrategies2019, bliksteinUsingLearningAnalytics2011, ciborowskaDetectingCharacterizingDeveloper2018, gaoExploringProgrammersAPI2020, liWhatHelpDevelopers2013}.

% NEW:
%  Instead of the Web, developers turned to their colleagues to acquire information, and such process is restricted to many factors, such as their availability and expertise. This cause issues in acquiring the information \cite{ko2007information}. 

\subsection{Impact on code}
The prevalence of code cloning \cite{hora2021, xiaWhatDevelopersSearch2017} has led researchers to investigate the degree to clones could propagate problematic code. 
Users of online code snippets report them to be outdated, incomplete, incorrect, poorly structured, verbose, lacking meaningful variable names and rationale, and indicate that code often does not address their problems  \cite{escobaravila2019, ragkhitwetsagulToxicCodeSnippets2019, treude_understanding_2017, wuHowDevelopersUtilize2019}. 
These problems have also been identified in data mining studies 
% of \StackOverflow{} \cite{ragkhitwetsagulToxicCodeSnippets2019, yang2016, zhang2018code, zhou2016} and programming tutorials \cite{nishi2019characterizing}. 
\cite{ragkhitwetsagulToxicCodeSnippets2019, yang2016, zhang2018code, zhou2016}, and observation and lab studies suggest that use of websites typically leads to poorer code \cite{acar2016you, baiExploringToolsStrategies2019}.
%Furthermore, such analyses have highlighted potential problems with licensing (\eg{} where code from a project under a particular license is posted to \StackOverflow{} with no indication of the associated licensing)~\cite{an2017stack, ragkhitwetsagulToxicCodeSnippets2019}.

% Propegation and Security
Data mining has also been a valuable tool for measuring propagation \cite{abdalkareem2017code, fischer2017stack, yang2017stack}. 
\etal{Abdalkareem} examined $1,496$ Android apps finding that $377$ had source in common with \StackOverflow{}, and the introduction of this code increased the number of subsequent bug fixes \cite{abdalkareem2017code}. 
Propagation of security vulnerabilities is a particular concern.
Lab study has shown that \StackOverflow{} users produce functional but insecure code \cite{acar2016you}, %unlike participants who use documentation who produce secure but not working code 
and data mining shows that a very high proportion of security-related snippets contain potential vulnerabilities \cite{fischer2017stack}, even in accepted answers, up-voted answers and answers from high reputation users \cite{meng2018secure}.
However, not all studies agree that cloning is problematic -- in a study of $1,244$ open-source projects, code reuse predicted neither the presence or absence of security vulnerabilities \cite{Gkortzis2021}. 

% Novices
Novice programmers may be particularly likely to generate and propagate problematic code. 
Static analysis of code written in an introductory programming exam identified both functional (syntax, logic) and stylistic errors \cite{delevStaticAnalysisSource2017}. 

% Research Gap
In this paper, we provide in-depth qualitative analysis of novice programmers' online information-seeking and code-cloning behaviours. 
Our research builds on knowledge from prior studies, using a combination of methods to provide a rich picture of online and in-editor activities, augmented with programmers' motivation and experiences of those activities, and the impact on code outputs.

\section{Methodology}
\begin{figure*}[tbhp]
\rule{0cm}{0cm}
\centering
% \includesvg[width=0.98\textwidth]{figures/methodology1.2}
\includegraphics[width=0.98\textwidth]{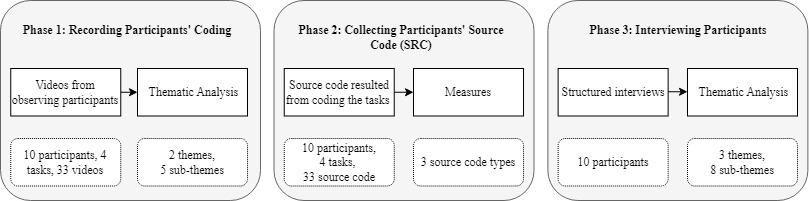}
\caption{Overview of research methodology: an online study in three sequential phases. A set of four video-captured programming tasks are followed by source code collection, and then an interview. The resulting qualitative data is thematically analysed.}
\label{fig:phases} 
\end{figure*}

We conducted an online study in three sequential phases: a set of four video-captured programming tasks, source code collection, and then an interview (see \hyperref[{fig:phases}]{Fig~\ref*{fig:phases}}). 
This approach allows us to triangulate multiple data sources to build a rich understanding of our participants' behaviour \cite{seamanQualitativeMethodsEmpirical1999}. 

The experiment was conducted with undergraduate students recruited from our university. %\UoM{}.
Prior to recruitment, a pilot was used to ensure the clarity and validity of tasks, and to assess overall experiment duration. Procedures for the experiment were reviewed and approved by the Ethics Committee at our university
%Department of Computer Science Ethics Committee at our university. %\UoM{}
% \footnote{Application reference: 2020-10022-16293.}. 
All data was anonymized at time of collection.

\subsection{Task design}
This study phase required participants to engage in a number of programming tasks appropriate to the participants' experience and expertise.
Having limited our recruitment to a single cohort of undergraduate students in their second year at our University, %\UoM{},
we could design our tasks to build on course materials from their first year of study. 
This included a course textbook~\JavaInTwoSemestersCitation{}, lecture slides, instructions for programming exercises, and sample solutions associated with those exercises. We extracted an initial set of thirty-three programming tasks from these materials.

To ensure task completion could be supported by online materials, specifically the \StackOverflow{} website\footnote{Prior research has indicated that \StackOverflow{} is a dominant resource for students and programmers when seeking help using websites}, we conducted a preliminary search of \StackOverflow{} using task keywords. 
Tasks with fewer than twenty \StackOverflow{} answers were removed from the candidate pool, leaving fourteen tasks for further consideration. 
% Examples of the code snippets available on \StackOverflow{} are presented in Figure~\ref{SO_Ava_T1and4}.
%
The remaining tasks were classified into three difficulty levels. 
Tasks that used only concepts/topics that were explicitly taught during the course unit lectures/exercises were classified as `easy` (n=5), those that used a combination of explicitly taught and additional material (\eg{} from the textbook) were classified as `medium' (n=5), and those that required substantial additional material were classified as `difficult` (n=4). 
From the classified tasks, we selected four tasks with a variety of difficulty levels: two easy, one medium and one difficult (See Table~\ref{tab:S3_tasks}).

% \begin{figure*}[tbp]
% \centering
% \includegraphics[width=\textwidth]{figures/stackoverflow_screenshots/Tasks1and4}
% \caption{An example of code snippets available on \StackOverflow{} for the first (left) and fourth (right) tasks.}
% \label{SO_Ava_T1and4}
% \end{figure*}

\begin{table*}[!t]
   \renewcommand{\arraystretch}{1.3}
   \footnotesize
   \caption{Study Tasks, Difficulty Levels, and Topics. Fixed Width Text Denotes Java Classes, Methods and Keywords. Topics Marked With an Asterisk (*) Were Not Explicitly Taught in the First Year Java Course, but Were Covered in  Additional Material.} 
  \label{tab:S3_tasks}
  \centering
  \begin{tabular}{l p{1.4cm} p{11cm} p{2.9cm}}
     \toprule \textbf{Task} & \textbf{Difficulty} & \textbf{Description} &  \textbf{Topics assessed} \\ \midrule
             1 \label{Task1} & Easy & Write a program that contains a two-dimensional array with the following values: 10, 20, 30, 40, 50, and 60. Declare and initialise a two-dimensional array with three rows and two columns. Then, using nested loops, print the array contents one by one in the same order. &  multi-dimensional arrays, nested loops\\
             2 \label{Task2} & Easy & Write and implement a program which converts a sum of money into a different currency. The user will enter the amount of money to be converted and the exchange rate. The program will contain separate methods for getting the sum of money from the user, getting the exchange rate from the user, calculating the conversion, and displaying the result. &  method definition, string input, string output\\
             3 \label{Task3} & Medium & Write a program to enter and confirm a suitable code name for a spy agent. Declare a \String{} variable and then get the user to enter a suitable name as a codename. Check that the codename meets the following requirements:
                \begin{enumerate}
                    \item The codename is greater than 6 characters in length.
                    \item The codename starts with the word “Agent”.
                    \item The codename ends with an “X” character. 
                \end{enumerate}
            If one of the conditions above was not met, print “INVALID CODENAME” and ask the user to re-enter a code name. &  \String{}, \startsWith{}*, \EndsWith{}*.\\
            4 \label{Task4} & Difficult & Using Java threading feature, create three threads where each has a unique name. Then, each thread should print the numbers from 1 to 100 in sequence order. For example, thread A will print numbers 1 to 100, thread B will print numbers 1 to 100, and thread C will print numbers 1 to 100. & threading*\\\bottomrule
       \end{tabular}
  \label{tab:addlabel}
\end{table*} 
\subsection{Study Procedure}
Our three-phase study took place over the \Zoom{} video conferencing software, and using the \Dropbox{} file hosting service.

\subsubsection{Preliminaries}
All participants were provided with a briefing sheet prior to participation (available as supplementary material\footnote{\label{fn:supplementary}\supplementary{}}).
A video call with each participant was initiated, and the researcher summarised the phases of the study. Once participants had a good understanding of the nature of their involvement, they were asked to provide verbal consent and recording of the call (audio and screen capture) commenced.

\subsubsection{Phase 1: Video-captured programming tasks}
At the beginning of this phase, participants were presented with a document summarising the four programming tasks. Participants were instructed that they should not overly concern themselves with the need to correctly solve all tasks, and that they could use any of their usual software, resources, or websites during task completion. 
Participants were then asked to share their screen and could begin solving the tasks using their preferred IDE or editor.

During the programming tasks, the researcher muted their mic and turned off their own video camera so as to minimise distraction. The researcher did not intervene or interrupt the coding tasks unless participants directly requested that they did so (\eg{} to help resolve a technical issues or task). The researcher did not provide any programming-related information that could help the participant solve tasks, even if requested to do so by the participant.

The programming task phase ended after $50$ minutes of elapsed time, or earlier if a participant indicated that all four tasks had been completed. 
Upon completion, screen sharing and recording was stopped, and the resulting video files (one per task) were saved for subsequent analysis.

\subsubsection{Phase 2: Source code collection}
Participants were then requested to share their final set of source code files for the four programming tasks. 
\Zoom{} chat was used to direct participants to a \Dropbox{} upload link.

\subsubsection{Phase 3: Interview}
The final phase was an audio-recorded structured interview. 
Participants were asked to provide their age, gender and describe their programming expertise/experience. They were asked about their experiences of completing the programming tasks, and the degree to which their behavior was representative of the participant's usual programming activity. Participants were additionally asked targeted questions about their use of websites during typical programming and during the experiment's programming task phase, code cloning behaviours during the programming task phase, and the relationship between website-based information retrieval and existing knowledge (\ie{} are they looking up things to refresh/jog their memory). The full questionset is provided in the supplementary material\textsuperscript{\ref{fn:supplementary}}.

\subsection{Participants and Recruitment}
Our target population were invited to participate through emails (sent to all second year computer science students) sent by %programme and 
course directors. 
Ten participants (five males and five females, aged $19-22$ years) responded and participated in the study (see Table \ref{tab:S3_demographic}). Participants were compensated with a £10 GBP Amazon gift certificate\footnote{Approximately \$14 USD.}.

\begin{table*}
\footnotesize
\centering
% \footnotesize
 \caption{Participants' Demographic and Expertise Responses. Experience is a Self-reported Value From 1 to 10, Where Novice=0, Intermediate=5, and Professional=10. Duration Refers to the Length of Time for Which They Have Programmed.}
 \label{tab:S3_demographic}
 \centering
  \begin{tabular}{lllllp{6.5cm}}
     \toprule \textbf{Participant} & \textbf{Gender} & \textbf{Age} & \textbf{Experience}  & \textbf{Duration} & \textbf{Programming languages} \\\midrule
            {P1} & {Male} & 20  & {5} & {Not available} & {\Python{}, \Java{}, \CSharp{}, JavaScript, and \RLang{}} \\
            {P2} & {Male} & 20  & {7} & {One semester} & {\Python{}, \Java{}, C, \CSharp{}} \\
            {P3} & {Female} & 19 & {5} & {One year} & {\Python{} and \Java{}} \\
            {P4} & {Female} & 19 & {5} & {5 years} & {4 programming languages} \\
            {P5} & {Male} & 20 & {5} & {8 years} & {\Python{}, \Java{}, JavaScript, \Typescript{}, \CSharp{}, \DotNET{}, C, C++, and \Swift{}.} \\
            {P6} & {Female} & 20  & {5} & {6 years} & {\Java{}, \Python{}, C, C++, Web development} \\
            {P7} & {Female} & 19  & {3} & {One year} & {\Python{}} \\
            {P8} & {Male} & 21  & {5} & {4 years} & {5 programming languages} \\
            {P9} & {Male} & 20  & {5} & {3 years} & {\Python{}, \Java{}, C++ and Pascal} \\
            {P10} & {Female} & 22  & {5} & {5 years} & {\Java{} along with $9$ programming languages} \\\bottomrule
       \end{tabular}
\end{table*}
\subsection{Analysis}
Interview recordings were transcribed for analysis. We analysed interview (phase 3) transcripts first, and subsequently used the knowledge gained to inform behavioral coding of the videos (phase 1) and source code (phase 2). Analysis was done by first author. %\Omar{}.
To ensure reliability of the coding, 30\% of the data (\ie{} three interview transcripts, ten videos, and source code from three participants) were additionally coded by third author%\Mohammed{}
, and any disagreements were resolved through a process of discussion and consensus,
with near-perfect agreement (Cohen's $k=0.83$, $k=0.87$, and $k=0.83$ for interviews, videos and source respectively; agreement $96\%$, $96\%$ and $93\%$).

Interview transcripts ($n=10$) were inductively coded using Braun and Clarke's six phases for reflexive thematic analysis \cite{braunUsingThematicAnalysis2006} and \NVivo{}.

Videos (screen recordings) were qualitatively analysed to observe \cite{seamanQualitativeMethodsEmpirical1999, stol2018abc} and quantify participants engagement with websites when solving the study tasks. 
The same researcher again followed a reflexive thematic analysis approach \cite{braunUsingThematicAnalysis2006} and linked the recorded videos to \NVivo{}, given the large size of the video files.

% NEW:
In the first step of the reflexive thematic analysis, the researcher watched the videos multiple times for content familiarisation. Then, the coding phase started with watching and coding the related behaviours, using a deductive approach. In contrast to the interviews, this approach was analyst-driven, using findings from our phase 3 interviews as a framework to identify relevant behaviors \cite{bakeman2012behavioral}. The choice of such an approach ensured behavioural coding of the videos that focused on the study's research questions and objectives, and helped the researcher to know what to expect when coding the videos. However, any interesting and related behaviours outside this framework were also captured to ensure unbiased data. To facilitate the coding, a marker was placed at the start of each task in the videos. Coding the videos was an iterative process involving many rounds, despite requiring more time and effort.

Initial coding observed behaviours based on the initial framework. Th subsequent round sought to identify related behaviours not covered by the framework. The third round was to extract behaviours related to the source code for later analysis. After this round, two criteria were established to ensure that the captured behaviours were meaningful and relevant. The criteria were that the behaviours had to be observed on the videos to constitute clear and reliable evidence and the behaviours should be measurable, in the sense that the videos provide enough information to support their serving as relevant evidence. The fourth round considered the mentioned criteria, checked the extracted behaviours and extracted any missing relevant behaviours. These multiple rounds ensured greater familiarity with the data and filtered behaviours down to relevant activities with multiple observed instances.

The next phase of the reflexive thematic analysis was grouping similar codes that form data patterns and share one overarching concept for the themes. For instance, codes that captured participants' behaviour toward online code adoption were grouped. The final set of themes was reviewed to ensure that no theme constituted a sub-theme to another theme or was not a theme in itself. Then, a thematic map was drawn to illustrate the final themes. Finally, a report on these behaviours was written.

% OLD:
% In contrast to interviews, video coding took a deductive approach using findings from our phase 3 interviews as a framework to identify relevant behaviors \cite{bakeman2012behavioral}. 

% Initial coding observed behaviours based on the initial framework. Subsequent rounds sought to identify related behaviours not covered by the framework, to generate hypotheses, and then to filter behaviours down to relevant activities with multiple observed instances. Codes were then grouped into themes and a thematic map developed. 

A total of $33$ videos (from a possible $40$) were analysed. Five participants did not attempt Task 4, and so no videos were captured for these participant-task pairings. A further two missing videos (\PTen{} Tasks 1 and \PTen{} Task 2) were attributed to a failure of the \Zoom{} recording facility. 

Source code produced by participants was analysed for the $33$ tasks for which video was captured. A mixture of quantitative and qualitative measures were established:
\begin{enumerate}
    \item \textbf{Quantitative: Compilation/Execution} -- Was the researcher able to compile and run the source code either using an IDE or Command line in Windows? 
    \item \textbf{Quantitative: Correctness} -- Does code execution produce the expected output (based on the specified tasks)?
    \item \textbf{Quantitative: Online Clones} -- the number of lines copied and pasted from online sources (final non-commented source lines only) retrieved from reviewing the videos.
    \item \textbf{Quantitative: Online Sources} -- the URLs from which cloned code was sourced (retrieved from reviewing the videos).
    \item \textbf{Qualitative: Clone Purpose} -- the perceived motivation for including cloned lines (retrieved from reviewing the videos). 
 \end{enumerate}

\section{Results} 

\subsection{The interview}
Our thematic analysis revealed three themes: reasons for website use, experiences using websites, and study specifications and issues.

% Tips:
% IT helps us understand the results. 
% Make sure all the related information is presented here. I used this as a basis for coding the behavioural coding. Thus, it need to reflect what I have done. 

% [I]
% [...]
% [Q:?]

\subsubsection{Theme 1: Usage of websites} %Previous name (was limiting the content): Reasons of websites usage.
Participants used websites for various purposes during their coding.

\paragraph{Websites as a reminder} \mbox{}\\
Nearly all participants reported that the websites, such as \StackOverflow{}, could serve as a reminder for information even if it were already known.
\begin{quote}
    I do not try to remember the code if I know where it is exactly. [...] I am just reminding myself from the sources. (\PThree{})
\end{quote}

%% \noindent Such reminding process comprised of information confirmation.
\begin{quote}
    [...] it is often the same thing looking up just to check if you remember it correctly. (\PSix{})
\end{quote}

%% \noindent One reason for difficulty in remembering programming information is the use of multiple programming languages, as mentioned by three participants.
% \begin{quote}
%      I usually use the website to refresh my memory because I just do not remember all of the intricacies of every single language because each of them have slightly different syntax [...]. \PTwo{}.
% \end{quote}
            
% \noindent Other reason that affect remembrance is the recency of coding.
% \begin{quote}
%      I have forgotten and not used it like for few months. \PNine{}.
% \end{quote}

\paragraph{Copy online code} \mbox{}\\
All participants reported copying and pasting online code, even the previously encountered code or problems, due to the easy-to-find online code. 
 \begin{quote}
    [Q: What about copying code that you have previously copied during websites search?]
    
    Yeah, I would copy it again; some code I know where I can find it, so whenever I cannot remember it by heart, I know exactly where to find and then just copied it and then it works. (\PEight{})
\end{quote} 
%% \begin{quote}
%     I look up the same thing every single time I program. \PFour{}
% \end{quote}
        
%% \noindent Most participants were confident that they acquired correct answers from the websites. On the other hand,
\noindent Most participants felt that the websites urged them to adopt available online content without checking, where the ease of obtaining online code gave a sense of no need to understand the code.
\begin{quote}
    [...] I went ahead and copied the first things I saw [...] I think this is very common from write code pull up a page get something from the internet and continue coding. (\PEight{})
\end{quote}
%% \begin{quote}
%          [Q: Do you spare the effort of memorise the information because it is available online?]
         
%          Yeah exactly, I know where to find it, so I do not want to learn it because it so easy to just pull up copy. \PEight{} 
% \end{quote}
    
%% \noindent The placement of syntax on the \Tutorial websites provided easy syntax extraction, as noted by three participants.
% \begin{quote}
%   [...] if I am looking for syntax, I usually go to  to those websites I mentioned before, and I am visually looking for these squares in different background, so I just filter what I see. If I see the bluish square with some code on it, I just look into it and usually I got what I need. \PTen{}.
% \end{quote}

\paragraph{Searching for information} \mbox{}\\
Participants described the procedures they took when searching the website for programming material. At the start, they get familiarised with the requirements to help determine the following step. 
\begin{quote}
   I think my approach would usually be the same, read though the questions a few times and give it a go and see where I feel it does struggling and then start seeing if there anyone else done it in a smart way or whatever. (\PNine{})
\end{quote}  

\noindent The following step is searching using the Google search engine without predetermining preference, as mentioned by most participants, and structuring the search query in a more appealing way related to the requested tasks.
\begin{quote}
    I did not use a particular website, just try to figure most concise way to word my question and just see what comes up. (\PNine{})
\end{quote}

%% \noindent When the results appeared, four participants followed three approaches to select the appropriate results. The first one was selecting of the most suitable website that deliver the required answers. 
% \begin{quote}
%       [...] I just used the search and choose the one which seemed the most appropriate. \PTwo{}
% \end{quote}
% \begin{quote}
%       [...] I chose whatever looks like it was answering my question, and most of it was fine. \PEight{}
% \end{quote}     

% \noindent The second one was choosing random websites from the Google search results, mostly the first one.
% \begin{quote}
%     [...] I used random websites as whatever came on the top of Google search that the first thing I checked out. \PFive{} 
% \end{quote}     

% \noindent The third approach was looking for easier answers.
% \begin{quote}
%       I just type my question and try to see from the description what answers are easier than go there. \PSeven{}
% \end{quote}            

\subsubsection{Theme 2: Websites experience}
This theme tackles programmers' websites' preferences and potential concerns.

\paragraph{Websites choices} \mbox{}\\
Half of the participants reported that \Tutorial websites provide clear, simple, understandable, self-sufficient, and easy to locate examples that offset the need to read the corresponding contents.  
\begin{quote}
     The one I usually chose is \Geeks or \WThree. These two because it explains the best what I need to find, and the answers are on the beginning, so I do not need to scroll and read through everything. (\PTen{})
\end{quote}
%% \begin{quote}
%     I think I used \Geeks at one point that usually quite useful because loads people post their like proper tutorials. (\PFive{})    
% \end{quote}
% \begin{quote}
%     \TutPoint and \WThree I generally use it a lot it because it straightforward and they provide pretty clear examples and do not rely too much on jargons and it is pretty easy to understand what you are looking for. (\Psix{})   
% \end{quote}

\noindent More than half of the participants mentioned using of \StackOverflow{} because it provides reliable and genuine examples aiding problem-solving.
\begin{quote}
     [...] I use \StackOverflow{} for for this one specific problem I was having because those questions are more specific. (\PSeven{})
\end{quote}
%% \begin{quote}
%     I chose SO because usually it is reliable. (\POne{})
% \end{quote}
 
\noindent The usage of the \StackOverflow{} is approached with care, as P5 noted:
\begin{quote}
      And there is \StackOverflow{}, but you need take that in with some caution sometime. (\PFive{})
\end{quote}

%% \noindent Other than websites' features, one participant mentioned that the more experienced programmers' suggestions determined her websites' selection.
% \begin{quote}
%     [Q: What the websites did you choose during the experiment? And why?]

%     Because it was recommended by the Uni stuff, and older generation. \PThree{}
% \end{quote}  

\paragraph{Issues with the websites} \mbox{}\\
Many participants reported issues when using the websites. Some online answers were functional, but they were not necessarily great. 
\begin{quote}
    [Q: When searching the websites, were you confident that you got the right answer?]
    
    Getting answers that works, but it is not always the best answers I would say. (\PFour{})
\end{quote}

\noindent Online answers additionally contributed to poor outcomes unsuitable for participants' intentions. 
\begin{quote}
     [Q: Did you find the websites easy to use, helpful, and correct?]
     
     One give me a misleading data, so I would say no. (\PEight{}) 
\end{quote}
            
\noindent Thus, participants reported the possibility of facing problematic code or incomplete code that misses important information. 
\begin{quote}
    % [Q: What did you mean using \StackOverflow{} with caution?]
    Sometimes the code on \StackOverflow{} can be a bit outdated, new version of framework or languages. Also, the users select an answer as the best answers that probably should not be, [...] like there are other answers better but not chosen by the user [...] who answer but it did not show up in the top. (\PFive{})
\end{quote} 
\begin{quote}
    The things it annoys me is website not showing the entire piece of code, so the thing that missing is something outside of the immediate bit of code they are showing. Like import statements, if the import statement is wrong and I cannot work out what it is, they very rarely show that I guess.  (\PFour{})
\end{quote}

\noindent Participants were uncertain about their ability to integrate online code due to the associated problems.  
\begin{quote}
    Some problems are sometimes you are not working with your own code, so you are do not know everything about it, you do not know how the code interact with the code you find online. (\POne{})
\end{quote}

\noindent Another impact was coding with unfamiliar programming aspects, where the websites did not help resolve such uncertainty.
 \begin{quote}
    [...] for the threading, I am not too familiar with threading, so I am not sure if it is what I wanted. (\PEight{}) 
\end{quote}

\subsubsection{Theme 3: Experiment's specifications and issues.}
This theme addresses participants' thoughts and concerns regarding aspects of the experiment.

\paragraph{Video recording} \mbox{}\\
More than half of the participants stated that their behaviours during the experiment followed their normal programming behaviours. 
\begin{quote}
    [Q: To what extends does the behaviours that you used today reflect the behaviours that you would normally use?]
    
    I think pretty accurately. (\PFour{}) %%, depending on how often I have been coding, recently I probably forgot fewer very simple things, but I guess, yeah I think so. (\PFour{})
\end{quote}

%% \noindent The online settings of the experiment is preferable than the lab-based because it relieves pressure from such experiment. 
% \begin{quote}
%     [Q: Do you feel me recording this experiment and being present affected your behaviour performance?]
    
%     Maybe if it was face-to-face, but since it virtual then no. (\PThree{})
% \end{quote}

\noindent Nevertheless, some behaviours were either not captured or did not reflect the programmers' normal behaviours. Two participants expressed that recording could conceal some behaviour or introduce unwanted factors to their programming activities.
\begin{quote}
    [...] I would probably open up my notes and some written notes which obviously that I cannot show on the screen. (\PTwo{}) %It is just a little pressure, it does not mean I will do something much different, it feels different, I am less relaxed. (\PTwo{}) 
\end{quote}
\begin{quote}
    [Q: To what extends does the behaviours that you used today reflect the behaviours that you would normally use?]
    
    It sorts of similar but in smaller scale. Usually I have got two monitors set up. (\PFive{}) %%so one monitors open with like problems descriptions and the websites that help me out, and write code on the other one so just happen in parallel. 
\end{quote} 
                
\paragraph{Time constraints} \mbox{}\\
Half of the participants felt that time constraints affected their performance in the experiment.
\begin{quote}
    So today I try do it relatively fast because of the time thing. Normally I will take it a bit slower and maybe read more things. (\PEight{})
\end{quote} 

\noindent The other half of the participants indicated no stress from the restricted time, and using some approaches would assist in regulating the time.
\begin{quote}
    What I do when I started problems, I know it is time constrain I sort try to get the first few really quick or the one I spot easier, so I try manage the time better, and I know I have done it so it does not much so time constrain does not stress me. (\PFive{}) 
\end{quote}

\paragraph{Task Difficulty} \mbox{}\\
Participants reported that the tasks were not difficult but required more searching.
\begin{quote}
    %% [...] I Google a lot more usually than I did today.
    % [Q: Why?]
    The tasks was not challenging. (\PFive{})
\end{quote}

\noindent However, four participants thought the fourth task could be challenging because they had no previous information, which may lead to not approaching the task.   
\begin{quote}
    %% [Q: What about the last task?]
    
    I have not got the time to read it [fourth task], but I am not familiar with threading, so that why I left it. (\PThree{})
\end{quote}

\noindent Two participants thought the problems were not from the task but from forgetting the information and time. 
\begin{quote}
     My Java was a bit rusty, slight, I did not remember a lot of things how to do in Java in particular. But no, I think if I have a bit more time I will finish them all.   (\PTwo{}) 
\end{quote}

\subsection{Behavioural coding}
%%Our thematic analysis of the recorded videos revealed two main themes concerning participants' strategies, including acquiring and utilising knowledge during development; these themes encompass seven sub-themes. Before presenting these themes, we describe the resources participants used along with their compiling activities. 

%%\subsubsection{Overview of the resources and compiling activities} 

% Comments from meetings:
    % providing the kind of here are the resources people used. Here, here's how they approach the tasks, here's what they got right and what they got wrong and this kind of thing
    % I guess it's keeping that view in your results section all the time on. What does this tell us about how people are using online resources? And that's the thing we're really interested in, and then you provide the additional information to fit around that so that people can understand what's going on
    
% Online resources & websites types
Participants used both Web-based and non Web-based resources to search.
All participants used websites at various stages. %%Most participants started their searches using the Google search engine and selected the first presented search results regardless of the website's name or the attached contents. Others chose the second website from the search results, which occurred in five instances by five participants. On the contrary, six participants in eight instances favoured a different approach by reading the returned search results and choosing accordingly. 
Most participants used \Tutorial{} websites to retrieve syntax, including \Geeks{}, \WThree{}, \JTpoint{}, \JSixSevenOne{}, \BegBook{} and \TutPoint{}. The choice of \Tutorial{} websites changed based on the tasks. %%as the first task involved \TwoDimensional{}, the second task multiple data types conversion, the third task \String{} functions, and the fourth task threading.
Other than \Tutorial{} websites, \StackOverflow{} was used for resolving errors, and blogs were used to aid for understanding.
%% Non online resources 
%In addition, participants used other than online resources during coding. For example, the tasks' file containing specified requirements available for participants to access during coding. In the first task, eight participants visited the tasks' file (\POne{}, \PTwo{}, \PThree{}, \PFour{}, \PFive{}, \PSix{}, \PSeven{}, \PEight{}), and most of their visits were commenced once, except for \PThree{} three times, \PSeven{} four times, and \PEight{} eight times. Similarly, all participants in the second task, except \PFour{} and \PTen{}, visited the tasks' file, and their visits were low, unlike \PThree{} and \PSix{} who visited that task's file six and ten times respectively. In the third task, nearly all the participants visited the tasks' file but not \PFive{} and \PNine. As compared to the other tasks, the third task possessed the most tasks' file visits as \PTwo{}, \PFour{}, and \PSix{} visited the tasks' file six, twice, and five times and \POne{}, \PThree{}, \PSeven{}, \PEight{}, and \PTen{} visited fourteen, fifteen, nine, thirteen, and seven times respectively. Lastly, the tasks' file visits in the fourth task were not high as \PFive{}, \PSix{}, \PSeven{}, \PEight{}, and \PTen{} visited three, five, four, four, and twice respectively. Other non-online resources used during developments were 
In addition, participants consulted the requirements document for tasks' requirements, course-related materials, previously written code, and the IDE. %%Specifically, \POne{} used coursework resources to copy the \Consolee{} syntax in the second task, \PFour{} used previous code to exploit syntax while coding the tasks like class declaring, \Main{}, \println{} statement and \Consolee{} syntax, and IDE provided syntax suggestions in the form of auto-complete used by \PTwo{} and \PSeven{} to get the \Main{} in the first and third tasks.
% Compiling events 
Participants edited and compiled their code, going through a cycle of observing and fixing errors post-compilation. %All participants ultimately complied code successfully for tasks one and two; all except \PNine did this for task three. All participants who attempted task 4 managed to compile their code for this. %% in a wide range of ways. All participants in the first task compiled their code and changed it according to the outcomes and the errors. \POne{} and \PTwo{} faced the most errors, such as logic and syntax errors from printing memory locations instead of array contents and missing brackets, causing extra time to resolve. All participants in the second task compiled their code and commenced fixing the produced errors, unlike \PFive{} and \PNine{}, who did not face any errors. \PTwo{}, \PFour{} and \PSix{} faced the most errors, such as run-time, syntax, and logic errors caused by missing syntax and incompatible data types, causing extra time to resolve. Compiling the code in the first and second tasks produced working code for all the participants. In the third task, all participants, except \PNine, compiled their code. \POne{}, \PThree{}, and \PEight{} faced a high volume of run-time, syntax, and logic errors, including errors related \String{} functions, missing syntax, and inappropriate use of \textit{``IF"} statements. They also spent considerable time resolving these errors more than other participants in both this task and other tasks. Although participants in the third task made numerous attempts to produce working code, more than half of the participants did not succeed in producing working code. During compiling the fourth task, \PSeven{} faced logic errors due to not printing the output of the threads' in order and run-time errors due to adopting unnecessary exceptions from online, causing extra time to resolve. Other participants either did not face any errors, such as \PFive{} and \PTen{}, or faced fewer errors like \PSix{} and \PEight{}. All participants who wrote a code on the fourth task produced working code. 

\subsubsection{Theme 1: Acquiring knowledge}
Participants searched websites for syntax and to clarify understanding. Searches yielded useful knowledge, but also caused problems.%% The following sub-themes will include details of the online searching sessions along with the encountered problems.

\paragraph{Looking for syntax} 
All participants searched websites for syntax either before or during coding. Before writing code, \PSix{}, \PSeven{}, \PEight{}, and \PTen{} in the fourth task and \PThree{} and \PEight{} in the first task searched for specific terms such as threading, \TwoDimensional{} and the \Main{}. 
The search for syntax while coding was noticed by nearly all participants who divided the tasks into meaningful searchable pieces and searched looking for syntax. Specifically, all participants in the first task, except \PTen{}, searched for \TwoDimensional{} and basic syntax like \For{}, \Main{} and array. In a similar manner, all participants in the second task, except \PTwo{} and \PTen{}, searched regarding converting data types and getting user inputs, such as \Scanner{} and \Consolee{}. \PFour{} and \PNine{} searched for basic syntax, including Java function structure and \println{}, similar to the previous task. All participants in the third task searched syntax to retrieve \String{} methods like \length{}, \startsWith{}, \EndsWith{}, \charAt{} and \substring{}. \PNine{} and \PSix{} searched for basic syntax, include \While{} loop \While{} loop and \AND symbol. In the fourth task, four participants (\PFive{}, \PSix{}, \PEight{}, \PTen{}) performed superficial searches, such as creating threads, assigning threads names and importing threads; \PEight{} used the whole question text to search at the task's start. 
 
\paragraph{Increase understanding}
Participants sought programming information related to the task along with searching for syntax. \POne{} and \PFour{} on the second task and \PSeven{} on the third task searched for syntax clarifications, such as the use of the \equals{} method for character type, the difference between \Float{} and \Double{} and the use of currency in Java. \PSeven{} and \PTen{} also searched to gather information about the threading feature in the fourth task.
 
%% \paragraph{Fixing errors}
% In order to resolve errors encountered during coding and compilation, participants resorted to the Web for assistance. During coding, \POne{} faced errors regarding non-correct syntax in the first and third tasks, such as \Main{}, \TwoDimensional{}, \startsWith{} and \equals{}. \PNine{} in the second task also encountered an error regarding the function structure and searched online for an example of a Java function. Regarding errors after code compiling, \PTwo{} and \PSeven{} faced errors in the first task caused by incorrect declaring of \TwoDimensional{} and \foreach{} loop. Five participants (\POne{}, \PTwo{}, \PFour{}, \PSix{}, \PSeven{}) in the second task faced errors, where \PFour{} encountered the most, and the errors were due to converting data types, missing elements such as using static in the method, issues with user inputs and misplacing data types caused by online suggestions. The third task has a few \String{} errors from \PFour{} on \length{} and \PTwo{} on declaring variables. Lastly, \PSeven{} and \PEight{} in the fourth task searched online to resolve errors on exceptions, confusion between threads and thread and thread names.
    
\paragraph{Experiencing problems during search} 
It can be observed that some online searches were unhelpful, causing additional time and effort. Eight participants struggled with online search across coding stages in all the tasks, and half of them returned to the previously visited Web pages. At the first task, \PTwo{} accessed various websites and searched for the nested \foreach{} loop, resulting in more search time and attempts, then eventually changing the chosen method. In another instance, \PTwo{} searched for \TwoDimensional{}, and the outcomes did not help succeed in printing array elements accurately but motivated further searches that also did not help fix the issue. Similarly, \PSeven{} searched repeatedly for a \TwoDimensional{} before writing it correctly. \PNine{} searched for the \Main{} and the \println{} statements without beneficial outcomes and resolved without searching. While coding the second task, \PFour{} struggled the most as she searched five times for \Float{} and \Double{}, truncate \Float{}, Decimal format and currency without valuable outcome; she sometimes returned to the previously searched results and chose another website. \PSix{} and \PSeven{} searched for \Scanner{} syntax, but their searches were not complete and caused further searches, and \PFour{} and \PNine{} conducted fruitless searches when converting data types from \String{} to \Float{}.  

Online searching issues continued in the third task. Five participants faced issues related to \String{}. \PTwo{} was affected the most as he searched for the \GetChar{} method in Java but adopted \String{} declaring instead, then continued searching for the equivalent of \GetChar{} from C++ in Java and accessed three websites without reaching an answer. The previously adopted \String{} declaring resulted in issues when compiling, causing two further search attempts, which was finally resolved by adopting a new code from websites; \PTen{} shared a similar struggle while searching for the \GetChar{} method. In addition, search instances conducted by \PThree{}, \PNine{} and \PTen{} caused further search attempts, such as searching for \String{} \length{} methods, split \String{} and \substring{}. \PEight{} searched for unnecessary syntax like \regex{} and \Matcher{}. Lastly, three participants (\PSeven{}, \PEight{}, \PTen{}) in the fourth task searched for how to start one thread after another and count threads, then faced unproductive content, causing them to reformulate their query and repeat the search.

\subsubsection{Theme 2: Utilising knowledge to write the code}
 % Needs elements of using knowldge that appear on the code. 
% Establishing the code, utilising knowledge to write the code 

Through seeking online knowledge, participants utilised the knowledge to develop their code. The following sub-themes will list participants' observed strategies when reusing online code. 
 
\paragraph{Exploiting search results to copy code}
All the participants copied online code snippets using two ways: a clipboard as a normal way and a visual way by looking at online code while writing it. There also appeared to be a further copy instance where participants examined particular code snippets from the websites and then wrote it in their code afterwards, and this copy instance will be labelled as a mental way. These three copy instances were conducted throughout the tasks with variance. Participants reused online code following the tasks' requirements. Seven participants in the first task, except \PFive{}, \PSix{} and \PTen{}, copied \TwoDimensional{}. Similarly, all participants in the second task, except \PTen{}, copied syntax, like declaring and importing \Scanner{} and converting syntax such as \String{} to \Float{} or \Integer{} or the other way around. In the third task, all participants, except \PFive{} and \PTen{}, copied \startsWith{}, \length{}, \EndsWith{}, \charAt{}, \substring{}, \Matcher{} and \Scanner{}. At last, all five participants copied online code, including a class header, printing outputs and declaring the variables. Participants within all the tasks copied basic syntax from online, such as \Main{}, \For{} and \println{} statements.

\paragraph{Implementing complicated approaches} 
Participants followed suggestions imposed by either websites or their code practices causing additional coding time and effort. Videos observed such approaches that did not necessarily appear in participants' source code. Five participants (\PTwo{}, \PThree{}, \PFour{}, \PFive{}, \PNine) in the first task followed online complex suggestions. In particular, \PTwo{} copied an online code that printed the array's contents using the \toString{} method, which caused printing memory places, but not the array contents. Three participants (\PThree{}, \PFour{}, \PFive{}) used online suggestions to manually assign values to the array without using a loop, and \PFive{} placed the array contents using three variables, each holding two array elements. \PTwo{} created four loops instead of two based on the online suggestions. A similar observation is valid for \POne{} but not from the online suggestions. In the second task, four participants (\PThree{}, \PSix{}, \PEight{}, \PNine) copied a user entry that used a \String{} data type from online, then converted it to \Integer{} data type. \PFive{} and \POne{} followed the same path without online suggestions. In the third task, six participants (\PTwo{}, \PThree{}, \PFour{}, \PFive{}, \PEight{}, \PTen{}) followed their code practices. \PTwo{} and \PThree{} faced issues declaring multiple \String{} variables and \PFour{} and \PTen{} made multiple unnecessary \IF{} statements that could be eliminated. \PEight{} chose \regex{} and \Matcher, which resulted in several searches, debugging and more time. As most of the code was copied online in the fourth task, the online suggestions were only problematic for \PSeven{}, who used three \RUN{} methods with different names and loops, causing more searches to write them correctly.

\subsection{Source code}
Table~\ref{tab:src_status} shows that nearly all participant-provided source code files compiled and executed successfully (28/33 files, 85\%), and just over half (19/33, 58\%) were considered to be correct (\ie{} they met the requirements described in the task). Across the four tasks, participants produced at least two executables (mean: $2.80$, median $2.50$) and one correct source file (mean: $1.88$, median $2.00$). Progressively fewer source code files were correct as the tasks progressed (max: $9$, min: $2$).

\begin{table*}
\centering
\caption{Results of Source Code Analysis.}
\label{tab:src_status}
  \begin{tabular}{l| llll |l} \toprule 
    & \textbf{Not Provided} & \textbf{Non-Executable} & \textbf{Incorrect} & \textbf{Correct} & \textbf{Online Clones} \\\midrule
    \textbf{Task 1} & 1: \PTen{} & 0 & 0 & 9: \POne{}-\PNine{} & 9\slash{}9 files: 1--5 lines\slash{}file\\
    & & & & & $\bar{x}$: $2.78$, $\tilde{x}$: $3.0$ \\
    \textbf{Task 2} & 1: \PTen{} & 0 & 4: \POne{}, \PThree{}, \PSix{}, \PEight{} & 5: \PTwo{}, \PFour{}, \PFive{}, \PSeven{}, \PNine{} & 9\slash{}9 files: 2--7 lines\slash{}file\\
    & & & & & $\bar{x}$: $4.67$, $\tilde{x}$: $5.0$ \\
    \textbf{Task 3} & 0 & 5: \PTwo{}-\PFive{}, \PNine{} & 2: \POne{}, \PSeven{} & 3: \PSix{}, \PEight{}, \PTen{}  & 9\slash{}10 files: 2--6 lines\slash{}file\\
    & & & & & $\bar{x}$: $3.80$ ($4.22$), $\tilde{x}$: $4.0$ ($4.0$) \\
    \textbf{Task 4} & 5: \POne{}-\PFour{}, \PNine{} & 0 & 3: \PFive{}, \PSix{}, \PTen{} & 2: \PSeven{}, \PEight{}  & 5\slash{}5 files: 6--24 lines\slash{}file\\
    & & & & & $\bar{x}$: $12.00$, $\tilde{x}$: $10.0$ \\\bottomrule
  \end{tabular}
  \begin{minipage}{17.5cm}
  \vspace{0.1cm}
  \small Note: In the First Four Columns, the Overall Number of Source Files in Each Category Is Followed By the List of Participants Whose Source Fell Into Each Category. Non-Executable Indicates That the Source Either Failed to Compile or Did Not Run to Completion. Incorrect Indicates That the Code Compiled and Executed Successfully but Produced Output That Was Not Compliant With The Requirements. In the Final Column, the Number of Files and Lines Containing Cloned Source Are Given, Together With the Mean ($\Bar{X}$) and Median ($\Tilde{X}$) Number of Cloned Lines Per File (Values in Brackets For Task 3 Indicate Averages for Files Containing Cloned Code).
  \end{minipage}
\end{table*}

All but one of the supplied source files (\PFive{} Task 3) contained cloned code, with each participant including an average of $2.67$ (mean, \POne{}) to $8.50$ (mean, \PSix{}) cloned lines per file (medians ranged from $1.5$, \PFive{}, to $6.0$, \PFour{} \& \PTen{}). 

%In the following, the status of each participant's source code will be discussed in detail and linked with participants' strategies from the videos along with a detailed online copy and paste activities retrieved from the video. 

\subsubsection{Correct} 
% Code status
There were instances of unnecessary syntax on the first and fourth tasks increasing the complexity of the correct source code. In the first task, \POne{} and \PTwo{} in the first task used four loop statements instead of two to solve the \TwoDimensional{} where \PTwo{} wrote such code based on online suggestions, \PFive{} copied online code and identified the \TwoDimensional{} using three array locations, and \PThree{} printed the locations of the array along with the contents. In the fourth task, \PSeven{} followed the online suggestions and included multiple \RUN{} methods for each thread, \For{}, and \println{} statements. 

% Linked with strategies 
Each of the correct source code were linked with video recordings to understand the strategies participants used to produce correct code. Regarding searching strategies, syntax searches were apparent throughout coding sessions for all participants, where they located items related to the task. Other than syntax searching, few searching occasions were attributed to clarifications. %%and resolving errors during coding and compiling. 
In addition, the online search resulted in multiple complicated issues faced by more than half of the participants who produced correct source code, which caused non-beneficial results and motivated further search attempts. Furthermore, all participants referenced the requirements document to check the requirements, and exhibited medium referencing in the first task, low in the second task, and high in the third task.

In terms of copying online code strategies, nearly all participants, who produced correct source code, copied online code in varied ways. %%, including normal, mental and visual, but the normal way is overwhelming. 
On the contrary, copying from non Web-based resources is a rare strategy done by one participant. %%Copied code were reused again from either previous or current tasks by half of the participants. 
In addition, participants in twelve correct code instances ran into complex issues, half of them from their choice and the other half from the online suggestions. %%At last, compiling the code and fixing the resulting errors caused six instances of high time spent fixing the errors.   

%% The instances of online copied code for the correct code type were analysed with the help of participants' video recordings (details of copy and paste activities can be found \ref{tab:online_usage}). The average of online copied code was calculated using the number of code copies for each participant divided by the number of participants. The first, second, and third tasks have lower online copy instances than the fourth task. Also, the \TwoDimensional{} with declaring, identifying, and assigning array values were the most online copied syntax because that the first task constituted the most of correct source code. Other syntax from other tasks were followed, such as \Scanner{}, \String{} related methods, and data types converting; \Main{} and \For{} were copied across the tasks. These syntax were retrieved mostly from tutorial websites.  

\subsubsection{Non-executable}  
Half of the third task participants (\PTwo{}, \PThree{}, \PFour{}, \PFive{}, \PNine{}) ($50\%$) failed to produce a functional compiled code. By analysing each source code within non-executable type, multiple reasons caused the code to be non-executable. Participants thought their code carried enough information while the code was incomplete, contained incorrect elements, or missed critical elements. In particular, \PTwo{} missed \SystemIn{} that caused incorrect \Scanner{}, used \return{} in the method while the method is void, used variable before declaring it, and wrote \texttt{else if} statement without including a condition. Similarly, \PThree{} used \texttt{else} without a condition and missed the \Scanner{} import, and \PTwo{} and \PThree{} copied from online incomplete \Scanner{} syntax and encountered difficulty when adopting the code. In addition, \PFour{} wrote four non-executable \IF{} statements, along with an incomplete \println{} statement. At last, the \PFive{} and \PNine{} source code contained incomplete \Main{} and missed a semicolon.

% Linked with strategies 
The strategies that lead to non-executable code were explored for a more in-depth understanding. Participants excessively searched for syntax during coding, except for \PFive{}, and other searches were for learning. %and resolving errors. 
Searching produce issues as \PTwo{} conducted a repeating search with no successful output, and \PThree{} and \PNine{} conducted unnecessary searches. Most participants referred to the requirements document moderately, except \PFive{} and \PNine{}. In the code strategies, online code was copied by all participants following mainly the visual way, except \PFive{}. Most participants made choices based on their practices that complicated their coding and used non Web-based code.%%, and built knowledge from second task code. In the end, compiling the code took average time except for \PThree{}, while \PNine{} did not debug the task. 

%% By analysing the instances of online copied copy for each participant in the third task (Table \ref{tab:online_usage} summarises the copying online code activates), the average copy is four instances, where they used tutorial websites to retrieve.

\subsubsection{Incorrect}  
Multiple reasons caused the source code to not comply with the specified requirements. In the second task, \POne{} and \PSix{} did not address \Float{} or \Double{} user entries (caused by copying online code for \POne{}) and \PThree{} and \PEight{} missed doing methods for each requirement. In the third task, the \String{} conditions wrote by \POne{} code did not function because it accepts non-string entries without processing, and \PSeven{} missed the \While{} loop for continuous user entry. In the fourth task, \PFive{}, \PSix{}, and \PTen{} printed threads but not in sequence order, and all solved the task engaging online. 

% Linked with strategies 
Participants in this source code type conducted some common behaviours. In syntax search, the online search was generally prevalent, where participants looked for tasks related to syntax. Facing complex issues were not high during searching activities. Nearly all the participants resorted to the requirements document with moderate to high access. For the code strategies, copying the code has normal observations where participants copy online code and faced some complex issues. %%One observation of the compiling process occurred to the P1 in the third task, where he spent 5 minutes resolving the errors. 

%% Regarding the instances of online copy details, the average of copying online code was five in the second task and two in the third task; however, high instances of copies are in the fourth task with an average of 15 instances. Regarding the copied code, copying the whole code was noticed in the fourth task to deliver the requirements (complete information visible in Table \ref{tab:online_usage}).

% \input{Inputs/04c_interview}

\section{Discussion}
This section discusses the study results and summarises the findings for each research question.

\subsection{RQ1: How do programmers use websites during programming?}
% Discussing searching activities
We found evidence from observations that the predominant use of websites was for retrieving syntax. %Such activities stress the importance of knowledge-seeking, similar to results from \cite{latozaMaintainingMentalModels2006, singerExaminationSoftwareEngineering2010, liWhatHelpDevelopers2013, ciborowskaDetectingCharacterizingDeveloper2018}. 
Participants in our experiment referred to various resources, including websites, the requirements document, course-related materials, previous code and the IDE; websites were used predominately to search for knowledge. Participants preferred referring to the \Tutorial{} websites to seek syntax, consistent with the findings of \cite{baiExploringToolsStrategies2019} but in contrast with other studies that suggest programmers use \StackOverflow{} for coding \cite{acar2016you}. They started their search by decomposing the tasks into searchable parts, similar to observations in \cite{sillitoManagingSoftwareChange2005}. 

% Discussing searching activities     
Two interesting observations were made in relation to searching: it was common to search for Java basic syntax such as \Main{}, \println{} statements and loops syntax, and previously taught syntax such as \TwoDimensional{} and user entry methods. %A possible explanation for the former search could be that participants disregard remembering the basic syntax and prefer to fetch them online. 
Interviews with participants showed they relied on websites to remind them of this basic information, which they may have been able to retrieve from memory if they had not had access to the websites. This tendency to search for basic syntax has also been seen in \cite{wangCharacterizingDeveloperBehavior2017, liWhatHelpDevelopers2013}. %The above reason could also affect the latter search (searching for the previously taught syntax) along with the duration between the experiment and the course taught to the participants. Fetching both syntax types from online could mean that 
Participants in our study preferred not to trust their memory of the syntax itself, but rather to rely on their ability to find it online, supporting the notion of using the websites as a reminder \cite{brandt2009two}. %The search instances matched those observed in \cite{wangCharacterizingDeveloperBehavior2017}, but 
%This study extends the work of \cite{wangCharacterizingDeveloperBehavior2017} by offering in-depth information that closely examines the behaviours and discovers difficulties. 
Participants also searched websites for, e.g., the equivalent of \GetChar{} from C++ in Java, suggesting that previous knowledge (especially knowledge of another programming language) plays a role in the search process. In some cases, participants searched continuously, without appearing to get relevant results. Failing to obtain relevant information in spite of repeated searchers has also been noted in \cite{acar2016you, wangCharacterizingDeveloperBehavior2017}. 

% Interesting and unfruitful searches. 
% Encountering issues during the online search could be explained by the fact that websites urge programmers to use the content without checking and encountering issues like outdated and incomplete code. Interviews support the notion of using online content without checking and facing problematic code. The preference for searching over examining search results and facing difficulties was also mentioned on \cite{starke2009searching} within the Eclipse platform, and facing code issues mentioned on \cite{ragkhitwetsagulToxicCodeSnippets2019}. While students have the nature of facing difficulties while coding, like producing errors \cite{delevStaticAnalysisSource2017}, Web usage influences their coding style and the information used. %Furthermore, the challenges faced when searching online, such as accessing not helpful Web pages, were similar to the findings from \cite{acar2016you, wangCharacterizingDeveloperBehavior2017}. It can be suggested that searching the websites during coding includes basic and familiar syntax and produces complexities such as unnecessary and unsuccessful searches. 

% Discussing code reusing activities 
Participants copied code from websites and used it within their programs. %%We designed tasks gradually from easy to hard, expecting that the more difficult task would reveal distinct findings.
Participants used code found online more in the fourth task than other tasks suggesting that the more difficult or unfamiliar the task, the greater the need to use online code. Using websites to help with unfamiliar tasks involves dealing with uncertainty about the information found, as expressed in the interviews. %However, this behaviour was also observed in the third, medium-difficulty task, which required understanding fewer novel concepts. %One possible explanation is that the fourth task contains more novel content than the third. 
Participants were observed to copy online code, including basic and taught syntax, in three ways: using copy and paste functionality; observing it and retyping it; memorising it then retyping it. They reported in the interviews that the reason for using online code is the ease with which it can be located. It appears easier to search for basic syntax every time it is required than to commit it to memory.% Reusing online code syntax suggests that programmers use websites to facilitate adopting code (that they probably used previously but did not retain internally) that helps their coding. %%Furthermore, an interesting observation is that similar syntax were used across the tasks without repeating the search, suggesting that programmers take advantage of their first initial search and not repeat the search; this observation is supported by participants in the interview reporting that they engaged in such constant copying of knowing code. The activity of repeatedly reusing similar syntax was in line with those using a semantic template \cite{kim2004ethnographic}, and the idea of syntax familiarity \cite{jacquesStudyingProgrammerBehaviour2021}. An observation of students' behaviours showed that students with low performance in a programming exam tend to refer to their old solutions more than top-performance students \cite{nygrenTrackingStudentsInternet2017}.

\subsection{RQ2: What are the effects of websites use on the resulting code?}

Using the websites while coding helped participants to produce correct code that met requirements. %Participants who produced correct code overcame online search issues and faced complicated approaches. This finding is similar to \cite{astromskisPatternsDevelopersBehaviour2017}, who showed that source code complexity implies more compiling and looking for online help. However, producing correct code is not always guaranteed. 
%Web impacted the code participants produced with instances of code that were either incorrect or non-executable. Performing online searches that were not helpful or unnecessary could produce non-executable code; this was also impacted by participants' practices of coding with complex approaches. 
Where participants produced non-executable code, this was not due to using websites per se, but rather poor programming practices. Nevertheless, there were many instances where participants did not exploit online content efficiently. %Interestingly, the instances of non-executable code were exhibited only on the third task. One possible explanation is that the third task is challenging, requiring more than other tasks. Web and the tasks' file access did not help reach code compliant with requirements. Participants referred to the tasks' file moderately to highly to check the requirements, but such access did not promise to deliver the code as per requirements specifications. The finding of this source code type is in agreement with \cite{acar2016you} findings which showed that using \StackOverflow{} makes participants produce functional but insecure code, while using official documentation produces secure but non-executable code. Participants had a free choice of accessing the websites to support their coding. They also had access to the tasks' file (for requirements checking). Overall, accessing the websites and tasks' files does not guarantee a correct code. In addition, the exploitation of the tasks' file did not ensure requirements fulfilment but rather the opposite. 
Copied code was sometimes neither suitable for nor required by the task. Examples are the manual assignment of array values, not printing array contents, unsuitable data types for user entries, redundant code and unnecessary multiple \RUN{} methods. Participants appeared to trust the code without giving it much scrutiny, at the expense of efficiency and effectiveness. It is not possible to know whether better code would have been written were the websites not available, but it is certainly the case that wholesale or unthinking inclusion of online code did not always result in satisfactory code. In the interviews participants reported that websites urged them to use code as presented, but we can see that this sometimes resulted in negative outcomes. The difficulty of reusing online code is also observed superficially in \cite{escobar-avilaSurveyOnlineLearning2019, xiaWhatDevelopersSearch2017}. The findings of this study suggest that while reusing online code can be effective, it can also cause complex problems that require time and effort to resolve. Programmers need to consider the associated time and effort while coding using the websites.

% Link with previous code issues.
Issues we observed in the code included: %The non-executable source code produced by participants as one of the written source code types is similar to \cite{abdalkareem2017code} who found that reusing code from \StackOverflow{} affected the functionality of the code but differed because the programmers wrote the code when using the websites. In addition, three problematic code were observed through the searching and coding strategies
 increased complexity; extraneous unwanted or unexpected outputs; and bugs \cite{zhang2018code, abdalkareem2017code}. Participants reported two additional issues in the interviews: outdated and incomplete problematic code \cite{yang2016, treude_understanding_2017, ragkhitwetsagulToxicCodeSnippets2019}. A possible explanation for these issues is that Websites state the code is safe to use as is; this was also mentioned in the interviews. Thus, it is necessary to reflect on the online presented content and ensure its accuracy.

% On the contrary, participants who produced non-executable code faced non-online complex code issues, referred moderately to the task file, copied online code visually, and produced non-fruitful online searches. Producing non-executable code could not be highly impacted by the websites but by the participants' practices. For example, participants neglected to include coding elements while they were presented online. %This view was supported by \cite{techapalokulNoviceProgrammersSoftware2017} who found that students programmers have the nature of introducing code smells in their programs. 

% Explanation other than strategies: 
Other factors may have affected the accuracy of source code, including task difficulty, time constraints and previous experience. While participants had previously been taught how to achieve most of the tasks, difficulty appeared to be a factor -- production of valid code reduced as task difficulty increased. However, participants reported in the interview that the tasks were not difficult to code, indicating they may have been unaware that code did not meet requirements. %We also expected (during the design of the tasks) that the websites would offset difficulties. In addition, participants noted in the interview that planning the experiment's time helps in coding the tasks. Furthermore, all the participants shared a similar background, with a few participants having more years of programming experience, such as P5, which did not reflect positively on the outputs and emphasised that the noticeable impacts occur regardless of the experience. 

\subsection{Recommendations}
The findings in this paper have applications for stakeholders across software engineering, 
including owners of the Websites, educators, researchers and tools builder. The following sections distill these findings into some concrete recommendations.

\subsubsection{Recommendations for owners of the Websites}
\begin{itemize}
    \item Owners of the websites should advise their content authors to consider the problematic code from the programmers' perspective and fully explain their online posted code.
    \item Owners of the websites should advise their content authors to consider that users with various experiences may consider their posted code.
\end{itemize}
 
\subsubsection{Recommendations for educators}
\begin{itemize}
    \item Educators should providing training in online information seeking, including how to search effectively, source selection, and appropriate expectations for online content.
    \item Educators should train students to engage in judicious code reuse, equipping them to make sound judgements about the suitability of code snippets. This includes supporting them in determining when (and which parts of) code snippets are relevant and recognising problematic code.
 \end{itemize}
 
\subsubsection{Recommendations for researchers}
\begin{itemize}
    \item Researchers can further investigate the search and code reuse strategies and propagation of problematic code by collecting or analysing ready sets of code. 
\end{itemize}
 
\subsubsection{Recommendations for tools builder}
\begin{itemize}
    \item Tools builder could exploit the findings by designing a tool that identify programmers' copy-and-paste activities when using websites and suggests follow-on activities such as the review of copied code for understanding, fit with requirements, and quality control.
\end{itemize}

\subsection{Limitation}

% NEW:
Participants' source code contained relatively few lines, providing little room to analyse the impact of website use. The study's design introduced no baseline with which to compare coding with or without using the websites. Asking participants to code without websites would not have been meaningful, and participants may have been reluctant to take part in a study designed that way. Also, repeating the study with the same participants would introduce previous exposure to the tasks. 

% Construct Validity:
In addition, to minimise threats to construct validity, a pilot study was used to validate and refine the tasks and the study's phases. One threat to construct validity could have been a concern about using appropriate tasks to reflect upon the coding activities. Task design ensures participant familiarity by using previous materials, and ensures online content support solving the tasks. Other threat may have emerged as a property of recruiting student participants from the University of Manchester. Whilst it was clearly indicated to participants that their data would be treated anonymously, to prevent impact on their study or outcomes, students may still have been reluctant to divulge behaviours that they thought academe would perceive negatively. Students at UK universities are regularly advised against activities that might constitute plagiarism, such as copying and pasting from external sources, and this therefore may have led students to minimise disclosure of these behaviours during their coding sessions and interviews. 

% Internal Validity:
This study does not set out to examine causal relationships, and the internal validity is of limited concern. However, potential influence by external factors could have included experiment time, settings and researcher availability influencing their coding.  
% Reliability
The study uses multiple data sources to ensure triangulation, including observations, source code and interviews. 
While the findings of this experiment resulted from solving programming tasks, the observed behaviours may vary based on the tasks. Time also play an important role in solving the tasks. In addition, steps were taken to involve multiple members of the research team at every step. An inter-rater reliability method was further conducted to increase the reliability of the findings.

\section{Conclusion}
% Present perfect for the study method like design, and past for the results. 

We conducted an online experiment to explore programmers' activities during coding tasks using websites, analysing the resulting source to uncover possible consequences on the code. Recordings and source code for ten programmers solving four programming tasks were collected, and participants were interviewed.  Recordings and interviews were thematically analysed, and source code analysed through a combination of quantitative and qualitative measures.

The observations have revealed that the vast majority of searches on websites were for syntax and involved tutorial websites. Syntax search comprised breaking down the task into searchable chunks and was usually for syntax that was basic, and had been previously taught. %Often participants did not find what they were looking for first time, and had to adjust and repeat their search. Additionally, the observations have found that reusing online code included known and basic syntax and involved 
Participants copied code in three ways: %normal, visual, and mental. 
using copy and paste functionality; observing it and retyping it; memorising it then retyping it. The copied syntax was not always appropriate or necessary for solving the task. %and was suggested by either online or participants' practices.

Participants produced source code in three categories: correct, which was working according to the requirements; non-executable; and incorrect, which was compiling/running but non-compliant with the requirements. 
% Participants' strategies during coding with the websites have been linked with their final source code. This association revealed that using the websites did not aid in averting writing code that is not working or not compliant with the requirements. In addition, producing code that was not compliant with the requirements was attributed to not understanding the requirements and highly referencing the requirements document. On the other hand, encountering complex issues when using the websites and understanding the requirements helped construct qualified code. 
Participants used various strategies during their coding using the websites, and these strategies were linked with participants' source code to investigate their outputs for any implications. Although coding with websites and encountering complex issues that increased task completion time and effort helped participants produce correct code,  using websites impacted the resulting code, producing either incorrect or non-executable code. Thus, using websites during coding produced incorrect code.
 
%The findings provided details of programming activities using the websites and how it impacts the produced code.
Programmers need to consider the time and effort it takes to use websites, reflect carefully on their requirements to help filter online content, and not presume that online content is accurate. Future work should explore these findings with other samples, for example, professional programmers.

\bibliographystyle{IEEEtran}
\bibliography{IEEEabrv,Study3}

\end{document}